\documentclass[review]{elsarticle}

\usepackage{lineno,hyperref}
\usepackage{float}
\usepackage{textcomp}
\usepackage{amsmath}
\usepackage{amsfonts}
\usepackage{bm}
\usepackage{xcolor}
\usepackage[ruled,vlined]{algorithm2e}
\usepackage{subcaption}
\usepackage{multirow}
\usepackage[bottom]{footmisc}

\journal{Aerospace Science and Technology}

\begin{document}
	
	\begin{frontmatter}
		
		\title{A Neural Process Approach for Probabilistic Reconstruction of No-Data Gaps in Lunar Digital Elevation Maps}
		
		% author
		\author[a1,a2]{Young-Jin Park}
		\ead{young.j.park@navercorp.com}
		
		\author[a2]{Han-Lim Choi \corref{cor1}}
		\ead{hanlimc@kaist.ac.kr}
		
		\cortext[cor1]{Corresponding author}
		\address[a1]{NAVER R\&D Center, NAVER Corp., Seongnam-si, Gyeonggi Province, South Korea}
		\address[a2]{Department of Aerospace Engineering, KAIST, Daejeon, South Korea}
		
		\begin{abstract}
			With the advent of NASA’s lunar reconnaissance orbiter (LRO), a large amount of high-resolution digital elevation maps (DEMs) have been constructed by using narrow-angle cameras (NACs) to characterize the Moon’s surface.
			However, NAC DEMs commonly contain no-data gaps (voids), which makes the map less reliable.
			To resolve the issue, this paper provides a deep-learning-based framework for the probabilistic reconstruction of no-data gaps in NAC DEMs. 
			The framework is built upon a state-of-the-art stochastic process model, attentive neural processes (ANP), and predicts the conditional distribution of elevation on the target coordinates (latitude and longitude) conditioned on the observed elevation data in nearby regions.
			Furthermore, this paper proposes \emph{sparse attentive neural processes (SANPs)} that not only reduces the linear computational complexity of the ANP $\mathcal{O}(N)$ to the constant complexity $\mathcal{O}(K)$ but enhance the reconstruction performance by preventing over-fitting and over-smoothing problems.
			The proposed method is evaluated on the Apollo 17 landing site (20.0$^{\circ}$N and 30.4$^{\circ}$E), demonstrating that the suggested approach successfully reconstructs no-data gaps with uncertainty analysis while preserving the high-resolution of original NAC DEMs.
		\end{abstract}
		
		\begin{keyword}
			Probabilistic Inference \sep Digital Elevation Models \sep No-Data Gap Filling \sep Sparse Approximation \sep Self-Supervised Learning \sep Neural Processes
		\end{keyword}
		
	\end{frontmatter}
	
	\section{Introduction}
	Lunar digital elevation models (DEMs) have been constructed with sensory data collected from the Lunar Reconnaissance Orbiter (LRO) \cite{chin2007lunar, smith2010lunar} and provided useful topographic information of the Moon since 2009.
	One of the most primary roles of LRO is to identify high-resolution elevations of the lunar surface by using data obtained from the Lunar Orbiter Laser Altimeter (LOLA) and the Lunar Reconnaissance Orbiter Camera (LROC) \cite{robinson2010lunar_a, robinson2010lunar_b}.
	LOLA provides a global topographic model of a resolution of 1024 pixel per degree ($\sim$30m at the equator) with high accuracy.
	On the other hand, LROC collects stereo observations with two narrow-angle cameras (NACs), which enables the generation of higher resolution ($\sim$5m at the equator) DEMs, called NAC DEMs \cite{tran2010generating}.
	
	Unfortunately, raw NAC DEMs contain regions of no-data gaps (voids) since stereo image matching processes often fail to match pixels near shadowed regions, breaking the continuity of the terrain map.
	Despite the problem, studies on no-data gap-filling algorithms for lunar NAC DEMs have not been sufficiently reported.
	Furthermore, a reconstruction for no-data gaps in lunar DEMs is not straightforward due to the following challenging issues:
	1) NAC DEM requires high-resolution reconstruction methodology,
	2) reconstruction algorithm must be reliable since it can affect related lunar studies or exploration missions, and
	3) NAC DEM is a large and high-resolution area map, thus a scalable approach should be applied.
	Accordingly, the main aim of this paper is to develop a scheme to robustly and efficiently inference no-data regions while maintaining the high-resolution of NAC DEMs.
	In particular, this paper adopts the deep-learning approach that has been gaining popularity in aerospace fields as well as artificial intelligence studies \cite{
		yang2016neural, chang2017adaptive, bagherzadeh2018nonlinear, furfaro2018deep, roy2019lunar}.
	
	One of the most conventional approaches for gap filling algorithms is to use an interpolation algorithm \cite{reuter2007evaluation, jassim2013image}.
	Interpolation based method primarily assumes spatial continuity of the map; the elevation value for a given point is more likely to be similar to the value of the nearby region than to that of the distant region.
	As such interpolation algorithms predict the unseen value as a weighted average of neighboring values where the weight is inversely proportional to the distance.
	Consequentially, interpolation approaches are difficult to account for regional characteristics and tend to focus too much on local information near the gap.
	Besides, they often provide blurred results and fail to resolve the first challenge.
	For more accurate and precise reconstruction, however, algorithms that adaptively reflect local characteristics and effectively aggregate information even away from the target region is needed.
	
	Alternatively, one can apply image inpainting algorithms by replacing the void filling task as image completion task; elevation maps can be viewed as single-channel images.
	Traditional inpainting algorithms \cite{bertalmio2000image, ballester2000filling, efros2001image, barnes2009patchmatch, komodakis2006image}, however, often fail to preserve details of surface information, thus are unsuitable for high-resolution reconstruction on large holes.
	On the other hand, data-driven approaches based on deep-learning methods such as a convolutional neural network (CNN) \cite{krizhevsky2012imagenet} and generative adversarial network (GAN) \cite{goodfellow2014generative, arjovsky2017wasserstein} have been recently studied \cite{iizuka2017globally, yeh2017semantic, yang2017high, gavriil2019void}.
	Although such approaches are known to generate unseen region with a very high-resolution, it is doubtful whether the predicted output from deep neural networks is reliable because they do not provide any uncertainty analysis.
	
\begin{figure}[t]
		\centering
		\includegraphics[width=1.0\textwidth]{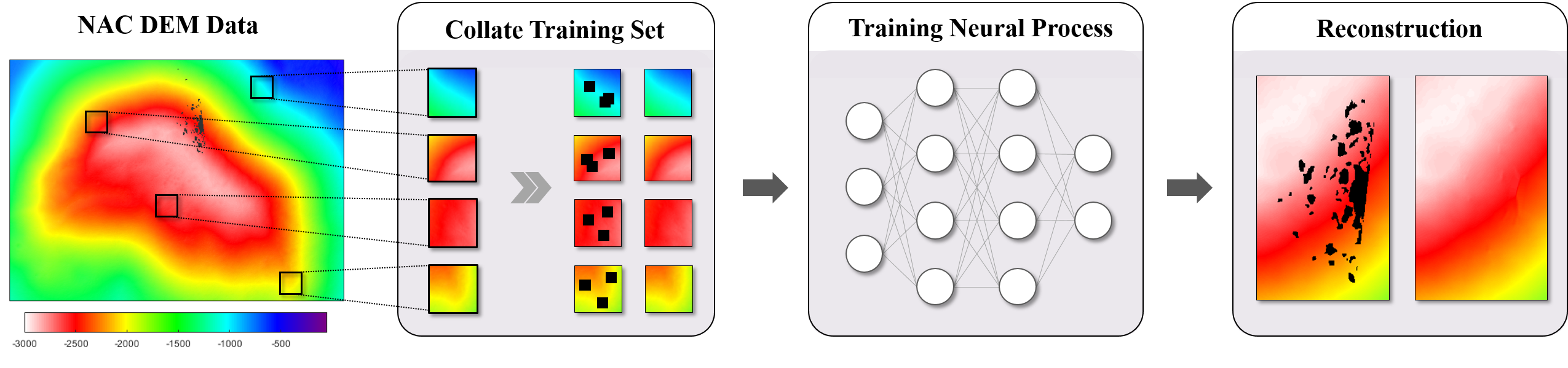}
		\caption{Flow diagram of the proposed neural process based no-data gap reconstruction method in Lunar NAC DEM.}
		\label{fig:arch}
	\end{figure}

	To resolve the presented challenges, this paper proposes a data-driven probabilistic inference scheme as illustrated in Figure \ref{fig:arch}.
	In particular, this paper primarily replaces the no-data gap reconstruction task as conditional inference problem.
	Then, we adopt the attention mechanism \cite{vaswani2017attention} based neural processes (NPs) \cite{garnelo2018neural, garnelo2018conditional} as a stochastic inference process to predict the elevation value on the shadowed region for given nearby observation data.
	The neural process model is trained in a self-supervised manner by using randomly masked NAC DEM data.
	We evaluated the effectiveness of our approach on the Apollo 17 landing site and showed that it fills no-data gaps of large-scale NAC DEMs in high precision and reliability. 
	
	%------------------------------------------------------------------------------
	
	\section{Preliminaries} \label{sec:bgd}
	Our method builds upon the attentive neural processes model \cite{kim2019attentive} which is a state of the art stochastic process model that incorporates attention approach into NPs.
	Here we briefly describe preliminary concepts about NPs and attention models.
	
	\subsection{Nomenclature}
	In the following sections, subscript $i$ and $j$ denotes $i^{th}$ and $j^{th}$ data set, respectively.
	Scalar, vector and matrix values are, represented using lowercase italic, lowercase bold italic, and capital bold italic, respectively.
	$\mathcal{N}(\mathbf{x}; \boldsymbol{\mu}, \mathbf{\Sigma})$ represents the probability density of vector $\mathbf{x}$ under the normal distribution with mean vector $\boldsymbol{\mu}$ and covariance matrix $\mathbf{\Sigma}$.
	$p(\mathbf{X}|\mathbf{Y};\mathbf{Z})$ denotes the probability of random variable $\mathbf{X}$ conditioning on random variable $\mathbf{Y}$ and parameter $\mathbf{Z}$.
	
	\subsection{Neural Processes} \label{sec:NP}
	The neural processes (NP) model is a flexible and powerful multi-dimensional regression model that approximates conditional distributions over functions given a set of observations \cite{garnelo2018conditional}.
	Consider a set of observation data consisting of $N$ input $\mathbf{x}_i \in \mathbb{R}^P$ and output $\mathbf{y}_i = f(\mathbf{x}_i) \in \mathbb{R}^Q$ pairs $\mathcal{O} = \{(\mathbf{x}_i, \mathbf{y}_i)\}_{i=1}^{N}$, and another set of $M$ unlabeled points $\mathbf{X}^* = \{\mathbf{x}_i^*\}_{i=1}^{M}$.
	The main aim of NP is to approximate a conditional distributions of $\mathbf{Y}^* = \{\mathbf{y}_{i}^* = f(\mathbf{x}_{i}^*)\}_{i=1}^{M}$ for unknown function $f: \mathbb{R}^P \rightarrow \mathbb{R}^Q$.
	Since it is analytically intractable to infer conditional probability for the nonlinear function $f$, NP introduces encoding process ($g_{\theta}$) that compresses each observation data into a finite-dimensional latent vector $\mathbf{r}_i \in \mathbb{R}^D$ and decoding process ($Q_{\phi}$) that reconstructs unseen data for a given data point and computed latent vectors:
	\begin{equation}
	p \left( \mathbf{Y}^* | \mathcal{O}, \mathbf{X}^* \right) \approx Q_{\phi}  \left( \mathbf{Y}^* | \mathbf{X}^*; \mathcal{R} \sim g_{\theta}(\mathcal{O}) \right) .
	\end{equation}
	where $\mathcal{R} =  \{ \mathbf{r}_i \}_{i=1}^{N}$ is a set of latent vectors while $\theta$ and $\phi$ are sets of learnable parameters for the encoding and decoding process respectively.
	
	For the decoding process, the mean-field approximation is used to achieve the scalable inference \footnote{The parallel inferences of $Q_{\phi}$ is enabled via GPU by mean-field factorization.}:
	\begin{equation}
	Q_{\phi} \left( \mathbf{Y}^* | \mathbf{X}^*; \mathcal{R} \right) = \prod_{\mathbf{x}^* \in \mathbf{X}^*} Q_{\phi} \left( \mathbf{y}^* | \mathbf{x}^*;  \mathcal{R} \right) .
	\end{equation}
	In more detail, $Q_{\phi}(\cdot)$ consists of mean and covariance functions $\mu_{\phi}(\cdot)$ and $\sigma_{\phi}(\cdot)$ parameterized by neural networks:
	\begin{align}
	Q_{\phi} \left( \mathbf{y}^* | \mathbf{x}^*; \mathcal{R} \right) &= \mathcal{N} \left( \mathbf{y}^* ; \mu_{\phi}(\mathbf{x}^*, \mathbf{r}^*), \sigma_{\phi}(\mathbf{x}^*, \mathbf{r}^*) \right) \label{eq:dec} \\ 
	\mathbf{r}^* &= aggregate \left( \mathcal{R} \right) \label{eq:agg}
	\end{align}
	where the $aggregate$ function is an arbitrary permutation-invariant function (e.g. sum, mean, max operation, etc.).
	The element-wise sum operation has been used for $aggregate$ function in previous works  \cite{garnelo2018neural, garnelo2018conditional}:
	\begin{equation}
	\mathbf{r} \equiv aggregate( \mathcal{R} ) = \mathbf{r}_1 \oplus \cdots \oplus \mathbf{r}_N \label{eq:sum_agg}
	\end{equation}
	
	Similarly, encoding process $g_{\theta}(\cdot)$ is also parameterized by neural network \footnote{The encoding process also can be considered as stochastic without loss of generality; but this paper focuses on deterministic process for the simplicity in implementation and operation.}:
	\begin{equation}
	\mathbf{r}_i = g_{\theta} \left( \mathbf{x}_i, \mathbf{y}_i \right) ~~~~~ \forall (\mathbf{x}_i, \mathbf{y}_i) \in \mathcal{O}  \label{eq:enc}
	\end{equation}
	
	Usage of the neural process has several big advantages as following.
	Primarily, NP is robust.
	Since NP is a probabilistic model, uncertainty analysis is available thus the predicted values are highly reliable.
	Secondly, NP is scalable.
	Note that Gaussian processes (GPs) \cite{rasmussen2006gaussian}, one of the most widely-used traditional stochastic processes, imposes an extensive computational cost of $\mathcal{O}(N^3)$, while NP can predict the data with a cost of $\mathcal{O}(N + M)$.
	Finally, NP can possess strong representation power by using the latest deep learning algorithms.
	
	\subsection{Attentive Neural Processes}
	In the early model of NPs, $aggregate$ function in \eqref{eq:agg} only take $\mathcal{R}$ as inputs and does not consider the inference point $\mathbf{x}^*$.
	Moreover, the encoding process in \eqref{eq:enc} does not take other data points into account.
	As such the conventional NPs infer latent variables $\mathcal{R}$ by merely using local information on each pixel, and therefore fail to compress the global information regarding whole observed data.
	As a result, overall processes could not have strong representation power and NP often suffers an under-fitting problem.
	To resolve the challenge, Kim et al.\cite{kim2019attentive} suggests attention based NP model, called attentive neural processes (ANPs), that incorporates the attention model into encoding and decoding processes.
	
	\subsubsection{Attention Models}
	Attention-based neural network architecture, the Transformer, is first introduced by Vaswani et al.\cite{vaswani2017attention}.
	After the invention of the Transformer, attention models have been applied to state of the art deep learning algorithms in the wide range of field (e.g. natural language processing \cite{devlin2018bert}, image processing \cite{yu2018generative}, recommendation \cite{zhou2018deep}, etc.), achieving superior performances.
	The key idea int the attention mechanism is to \textit{learn} to adaptively determine the associations among data: it automatically figures out the more and less important data points (regions) to pay attention.
	Consider a lunar crater classification task, for instance, while vanilla neural process places equal importance on every pixel data, attention models try to focus on pixels in crater regions automatically.
	One of the biggest advantages of using attention model is that the representation power of the model drastically increases.
	
	Consider a set of key-value (input-output) pairs $\{ \mathbf{k}_i, \mathbf{v}_i \}_{i=1}^N$ and queries (target) $\{ \mathbf{q}_i \}_{i=1}^M$.
	We can build a key matrix $\mathbf{K}  \in \mathbb{R}^{M \times P} $, a value matrix $\mathbf{V} \in \mathbb{R}^{N \times Q}$, and a query matrix $\mathbf{Q} \in \mathbb{R}^{N \times P}$ by stacking data along the column:
	\begin{equation}
	\mathbf{K} = \Big[ \mathbf{k}_1 , \cdots, \mathbf{k}_N \Big]  , ~~ 
	\mathbf{V} = \Big[ \mathbf{v}_1 , \cdots, \mathbf{v}_N \Big]  , ~~ 
	\mathbf{Q} = \Big[ \mathbf{q}_1 , \cdots, \mathbf{q}_M \Big] .
	\end{equation}
	Scaled dot-product (SDP) attention function is given by:
	\begin{equation}
	SDP(\mathbf{Q}, \mathbf{K}, \mathbf{V}) \equiv \mbox{softmax} \Big( \frac{\mathbf{Q} \mathbf{K}^T}{\sqrt{P}} \Big) \mathbf{V} . 
	\label{eq:sdp}
	\end{equation}
	To be specific, SDP function predicts $j^{th}$ output for a given query $\mathbf{q}_j$ as a weighted average on the set of observation values $\{ \mathbf{v}_i \}_{i=1}^N$ by giving higher weight on the value $\mathbf{v}_i$ that has a larger dot product of key $\mathbf{k}_i$ and $\mathbf{q}_j$.
	
	Vanilla SDP function does not have any trainable parameters.
	To enhance the representation power of attention network, linear transformation layers are often preceded before SDP function:
	\begin{equation}
	Attention \big( \mathbf{Q}, \mathbf{K}, \mathbf{V} \big) \equiv SDP \big( \mathbf{Q} \mathbf{W}^Q, \mathbf{K}  \mathbf{W}^K, \mathbf{V} \mathbf{W}^V \big)
	\end{equation}
	where $\mathbf{W}^Q, \mathbf{W}^K \in \mathbb{R}^{P \times D_q}$ and $\mathbf{W}^V \in \mathbb{R}^{P \times Q}$ are query, key, and value embedding matrices respectively.
	Note that using additional linear layers also allows the attention function to take into account the heterogeneity of query, key and value matrices.
	
	Furthermore, we can extend the attention model to a multi-headed model to consider $K$ number of relational processes:
	\begin{equation}
	MA^K \big( \mathbf{Q}, \mathbf{K}, \mathbf{V} \big) \equiv \mathbf{w}_0 + \sum_{k=1}^{K} Attention^k \big( \mathbf{Q}, \mathbf{K}, \mathbf{V} \big)\mathbf{W}^k
	\end{equation}
	where $Attention^k$ is $k$-th attention function, $\mathbf{W}^k$ are corresponding embedding matrix, and $\mathbf{w}^0$ is bias vector.
	$K$ is user defined hyperparameter.
	
	\subsubsection{Attentive Neural Processes} \label{sec:anp}
\begin{figure}[t]
		\begin{subfigure}{.44\textwidth}
			\centering
			% include first image
			\includegraphics[width=0.9\linewidth]{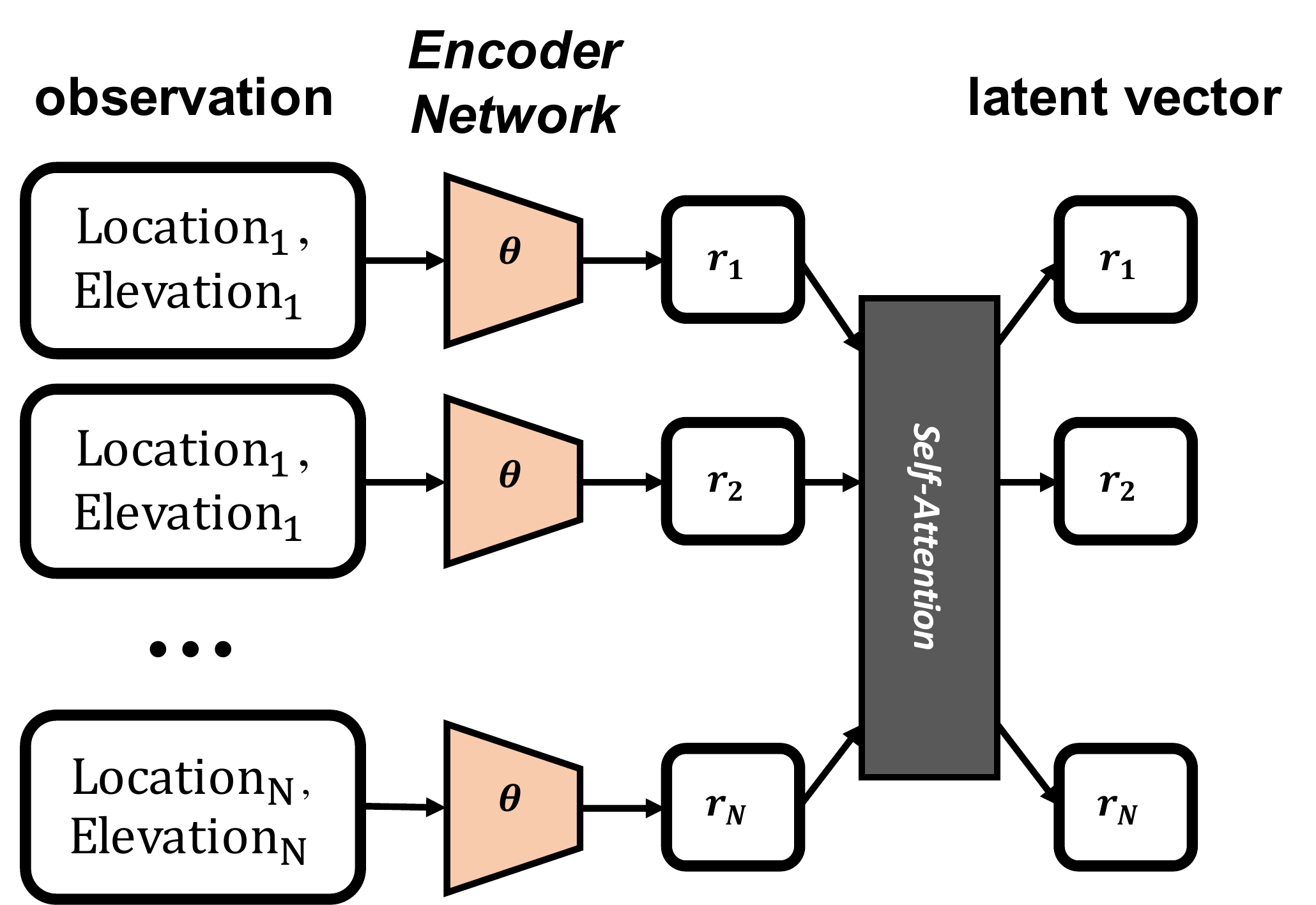}  
			\caption{Encoding Process}
			\label{fig:sub-first}
		\end{subfigure}
		\begin{subfigure}{.55\textwidth}
			\centering
			% include second image
			\includegraphics[width=0.9\linewidth]{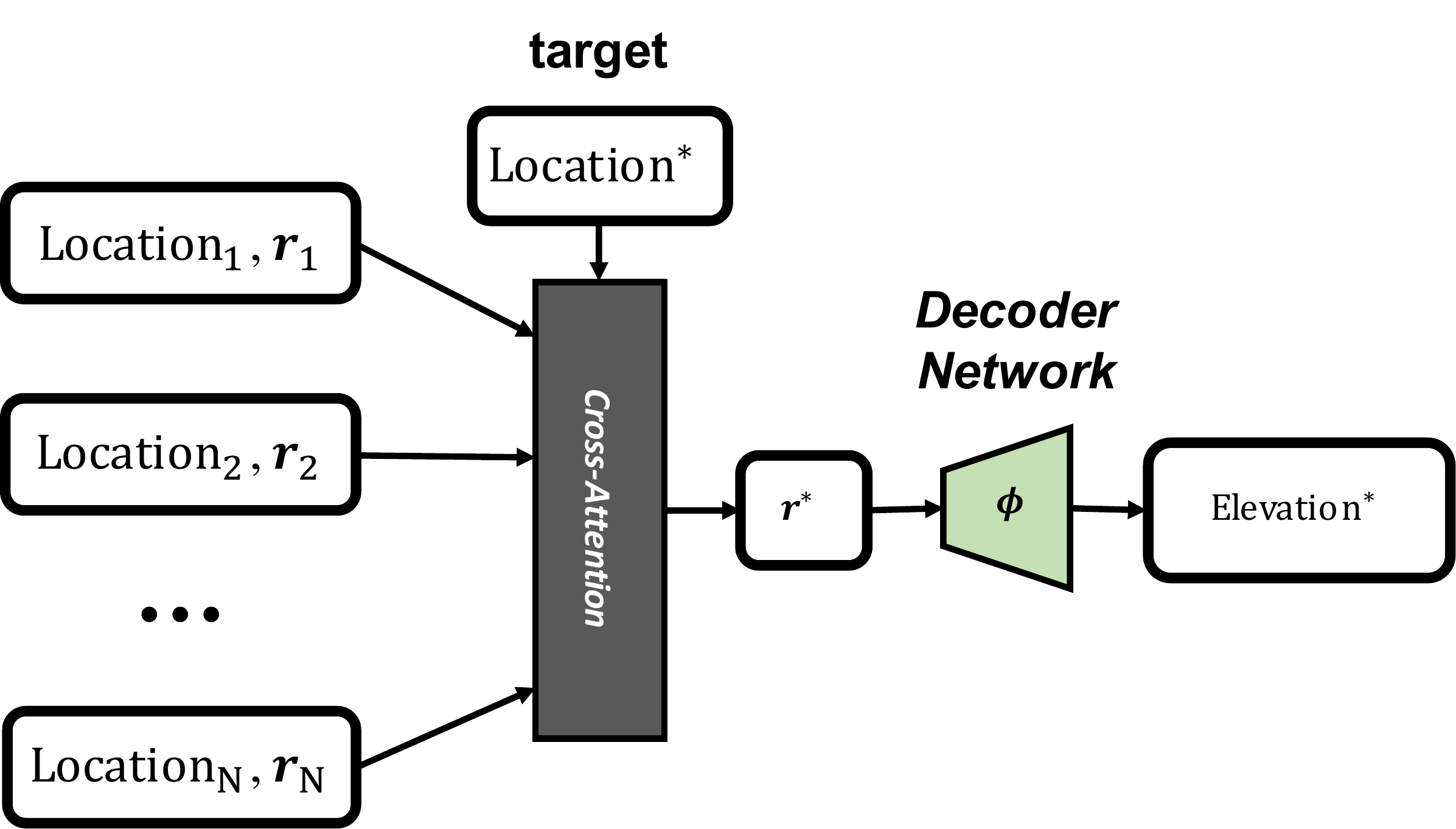}  
			\caption{Decoding Process}
			\label{fig:sub-second}
		\end{subfigure}
		\caption{The architecture of attentive neural process}
		\label{fig:anp}
	\end{figure}

	The primary difference between NP and ANP is that ANP adopts multi-headed \textit{self-attention} (attention where keys and queries are identical) on the encoding processes:
	\begin{equation}
	\mathbf{R}_{attn} = SA \left( \mathbf{R} \right) = MA^{K_e} \left( \mathbf{R}, \mathbf{R}, \mathbf{R} \right)
	\end{equation}
	where $\mathbf{R} = \left[ \mathbf{r}_1 , \cdots, \mathbf{r}_N \right]$ is a matrix of latent vectors.
	ANP additionally adopts multi-headed \textit{cross-attention} instead of $aggregate$ function in equation \eqref{eq:agg}:
	\begin{equation}
	\mathbf{r}_i^* = MA^{K_d} \left( \mathbf{x}_i^*, \mathbf{X}, \mathbf{R}_{attn} \right)
	\end{equation}
	where $\mathbf{X} = \left[ \mathbf{x}_1 , \cdots, \mathbf{x}_N \right]$ is a matrix of input vectors.
	The graphical representation of ANP architecture is illustrated in Figure \ref{fig:anp}.
	
	%---------------------------------------------------------------------------------
	\section{Probabilistic Reconstruction Framework for No-Data Gap in NAC DEMs} \label{sec:methods}
	\subsection{Problem Formulation}
	Primarily, we formulate the no-data gap reconstruction problem as a probabilistic inference problem.
	Denote the $L$ latitude-longitude pairs on the non-shadowed regions of NAC DEM, $\mathcal{D} = \{\mathbf{x}_i=(\lambda_i, \varphi_i)\}_{i=1}^{L}$ and the elevation for each point $\{\mathbf{y}_i\}_{i=1}^{L}$.
	Similarly, we denote the location and elevation for shadowed region $\{\mathbf{x}^*_i\}_{i=1}^{T}$ and$\{\mathbf{y}^*_i\}_{i=1}^{T}$, respectively.
	The goal of no-data gap reconstruction is to compute the probability distribution: $q \big( \mathbf{y}^* \big) = p \big( \mathbf{y}^* | \mathbf{x}^*, \{\mathbf{x}_i, \mathbf{y}_i\}_{i=1}^{L} \big)$.
	However, it is computationally extensive to consider every elevation data of high-resolution NAC DEM.
	Instead, we approximate the distribution as:
	\begin{align}
	& q \big( \mathbf{y}^* \big) \approx p \big( \mathbf{y}^* | \mathbf{x}^*, \{\mathbf{x}_i, \mathbf{y}_i\}_{i \in \mathcal{W}(\mathbf{x}^*)} \big) \\ \nonumber
	& \mathcal{W}\big( \mathbf{x} =(\lambda, \varphi) \big) \\ \nonumber
	& ~~~~ = \{ i \mid ~ |\lambda_i - \lambda| \le \frac{1}{2} w_\lambda, ~ |\varphi_i - \varphi| \le \frac{1}{2} w_\varphi \}
	\end{align}
	where $w_\lambda$ and $w_\varphi$ are user-defined window sizes.
	As an inference model, attentive neural process model, $Q_\phi(\cdot)$, is used to estimate the posterior distribution of elevation value for given target points.
	In order to normalize the range of input values for effective computation, we used a relative location to the target point rather as NP inputs.
	Note that since the encoding and decoding processes are both parameterized by the neural network in an amortized manner \cite{kingma2013auto, gershman2014amortized} as illustrated in equation \eqref{eq:dec} and \eqref{eq:enc}, we utilized only a single neural process model that can be applied to an arbitrary target region.
	
	\subsection{Model Architecture}
	\subsubsection{Encoder}
	We built the encoder network with three neural-networks: location encoder, elevation encoder, and latent encoder.
	Location encoder ($g_{l}$) and elevation ($g_{e}$) encoder linear transforms the location ($\mathbf{x}_i$) and elevation ($\mathbf{y}_i$) information into the latent space with dimension $D$:
	\begin{align}
	g_{l}(\mathbf{x}_i) & \equiv \mathbf{x}'_i = \mathbf{W}_l\mathbf{x}_i \\ \nonumber
	g_{e}(\mathbf{y}_i) & \equiv \mathbf{y}'_i = \mathbf{W}_e\mathbf{y}_i
	\end{align}
	where $\mathbf{W}_l \in \mathcal{R}^{D \times 2}$ and $\mathbf{W}_e \in \mathcal{R}^{D \times 1}$ are learnable parameters.
	Then latent encoder ($g_{latent}$), the neural network ($\mbox{mlp}_e$) that consists of two fully connected layers with 1024 hidden units and ReLU activation, aggregates the transformed location and elevation information into the latent vector $\mathbf{r}_i$ with the same dimension $D$:
	\begin{align}
	g_{latent}(\mathbf{x}'_i, ~ \mathbf{y}'_i) \equiv \mathbf{r}_i = \mbox{mlp}_e(\mathbf{x}'_i + \mathbf{y}'_i)
	\end{align}.
	Finally, as presented in Section \ref{sec:anp}, we aggregate latent vectors by using the 2-headed self-attention layer to compute $\mathbf{R}_{attn} \equiv \{\mathbf{r}'_i \}_{i=1}^N$.
	
	\subsubsection{Decoder}
	Decoder network consists of a 2-headed cross-attention layer and one neural network.
	Cross-attention layer estimates the latent vector for the target location $\mathbf{r}^*$ with $\{\mathbf{x}'_i\}_{i=1}^N$ as key, $\{\mathbf{r}_i'\}_{i=1}^N$ as value, and the transformed target location $g_l(\mathbf{x}^*)$ as query.
	Finally, the neural network ($\mbox{mlp}_d$) that consists of two fully connected layers with 1024 hidden units and ReLU activation, decodes the latent vector into the elevation $\mathbf{y}^*$:
	\begin{align}
	\mathbf{y}^* = \mbox{mlp}_d(\mathbf{r}^*)
	\end{align}
	
	\subsection{Sparse Neural Processes}
	Although the computational cost of ANP, $\mathcal{O}(ND^2)$, is much lighter than the one of Gaussian processes, the dot-product operation in \eqref{eq:sdp} is still a computationally and memory extensive.
	As a result, the vanilla ANP algorithm cannot be applied to large window sizes.
	Furthermore, using large amounts of information does not always lead to better results since the model can be easily over-fitted and leads the slow convergence.
	As such, we propose \emph{sparse attentive neural processes} (SANPs), a self-supervised learning scheme that not only is applicable to the large scale window size but enhances the model performance.
	\footnote{In the same way, neural processes are extended to sparse neural processes (SNPs)}.
	
	For each training iteration, we randomly select $B < L$ numbers of data points $\mathcal{D}'$ among $\mathcal{D}$.
	For each data point $\mathbf{x}_i$ in $\mathcal{D}'$, we collect the DEM data of window size $\mathcal{S}_i = \{(\mathbf{x}_j, \mathbf{y}_j)\}_{j \in \mathcal{W}(\mathbf{x}_i)}$.
	After this, we randomly sample $K \ll |\mathcal{S}_i|$ context points $\mathcal{O}_i$ for each $\mathcal{S}_i$ under the probability distribution:
	\begin{equation}
	p_{sample}(\mathbf{x}; \mathbf{x}_i)=\begin{cases}
	\exp(-\frac{1}{\alpha} || \mathbf{x} - \mathbf{x}_i||_2), & \text{if $\mathbf{x} \ne \mathbf{x}_i$}.\\
	0, & \text{otherwise}.
	\end{cases}
	\end{equation}
	where $\alpha$ is a user-defined sampling temperature.
	Note that the context points are uniformly sampled for $\alpha=\infty$ while the top-$K$ closest points are selected for $\alpha=0$.
	Finally, the model is trained to maximize the probability of elevation values for given sampled context points:
	\begin{equation}
	J = \frac{1}{B} \sum_{i=1}^{B} \Big[ Q_\phi(\mathbf{y}_i \mid \mathbf{x}_i, g_\theta(\mathcal{O}_i)) \Big]
	\end{equation}
	We used Adam \cite{kingma2014adam} algorithm for the gradient descent optimizer.
	The overall algorithm flow is described in Algorithm \ref{alg:dem}.
	
	\subsection{Data Augmentation}
	Data augmentation technique has been widely applied in various machine learning models, especially in computer vision fields, which effectively enhance the model performance.
	Data augmentation reduces generalization errors, particularly when the number of data is not sufficiently secured, such as the DEM problem.
	Therefore, this paper applied two data augmentation techniques, rotation and scaling, as described in Algorithm \ref{alg:da}.
	
	\begin{algorithm}[t]
		\SetAlgoLined
		Initialize dataset $\mathcal{D}$  \\
		Initialize ANP paraemters $\theta, \phi$  \\
		Initialize learning rate $\eta$ \\
		\While{Converged}{
			$J$ = 0 \\
			Sample $B$ data points $\mathcal{D}'$ from $\mathcal{D}$ \\
			\For{$\{ \mathbf{x}_i, \mathbf{y}_i \}$ in $\mathcal{D}'$}{
				Gather DEM data $\mathcal{S}_i$ near $\mathbf{x}_i$ \\
				Sample context points $\mathcal{O}_i$\\
				Apply data augmentation to $\mathcal{O}_i$ \\
				Collate training triplet $\{ \mathcal{O}_i, \mathbf{x}_i^*, \mathbf{y}_i\}$ \\
				Compute the log-likelihood of the elevation at target regions: $p = Q_\phi(\mathbf{y}_i | \mathbf{x}_i, g_\theta(\mathcal{O}_i))$ \\ 
				$J = J + \frac{1}{B} p$
			}
			Train $\theta$ and $\phi$ with gradient-descent algorithm
		}
		\caption{Sparse Attentive Neural Processes}
		\label{alg:dem}
	\end{algorithm}
	
	\begin{algorithm}[t]
		\SetAlgoLined
		\textbf{Input}: context points $\mathcal{O}_i$ \\
		Sample rotation angle $\theta_i$ in $\left[ 0, 2\pi \right]$ \\
		Sample scaling factor $s_i$ in $\left[ 0.5, 1.5 \right]$ \\
		\For{$\{\mathbf{x}_j=(\lambda_j, \varphi_j), \mathbf{y}_j \}$ in $\mathcal{O}_i$}{
			$\lambda_j \leftarrow \lambda_j \cos (\theta_i) - \varphi_j \sin (\theta_i) $ \\
			$\varphi_j \leftarrow \lambda_j \sin (\theta_i) + \varphi_j \cos (\theta_i) $ \\
			$\mathbf{y}_j \leftarrow \mathbf{y}_j \times s_i$ \\
		}
		\caption{Data Augmentation}
		\label{alg:da}
	\end{algorithm}
	
	%---------------------------------------------------------------------------------
	\section{Experiments}
	To demonstrate the efficiency of the proposed method, we compared the reconstruction performance with baseline methods on NAC DEM data at the Apollo 17 landing site (20.0$^{\circ}$N and 30.4$^{\circ}$E)
	\footnote{The data is collected from \hyperlink{http://wms.lroc.asu.edu/lroc}{http://wms.lroc.asu.edu/lroc}}.
	Data consists of 4.5M points (pixels) of size 0.005 $km$ * 0.005 $km$.
	We used 10k points for valid and test data each, before training models.
	As expected, valid and test data is not used for training procedures.
	Our model is implemented in \emph{PyTorch} \cite{NEURIPS2019_9015} on a single Nvidia V100 GPU, while the baseline algorithms are implemented in \emph{SciPy} \cite{jones2001scipy} and \emph{OpenCV-Python} \cite{opencv_library}.
	The code will be available after publication.
	
	\subsection{Model Specification}
	For sparse neural process models, window sizes $w_\phi$ and $w_\varphi$ are set to 0.5 $km$ ($\sim$ 100 pixels) so that they can adequately cover the size of the no-data gap.
	Hyperparameters, sampling size ($K$), sampling temperature ($\alpha$), and latent dimensions ($D$), are set to 100, 0.4 $km$, and 512 respectively through ablation studies in section \ref{sec:abl}. For the vanilla neural process models, the window sizes are adjusted to 0.05 $km$ ($\sim$ 10 pixels) to match the number of context points, and the latent dimension was equally set to 512.
	Furthermore, we compare our model with the following baselines as well as neural process models:
	\begin{itemize}
		\item \textbf{Linear} is an extension of linear interpolation for 2 - dimensional data, using the closest $2 \times 2$ pixels.
		\item \textbf{Cubic} is an extension of cubic interpolation for 2 - dimensional data, using the closest $4 \times 4$ pixels.
		\item \textbf{Nearest} fills with the value of the nearest point.
		\item \textbf{Navier Stokes} \cite{bertalmio2001navier} is one of the most well-known image inpainting algorithms based on fluid dynamics utilizing partial differential equations. 
		\item \textbf{Telea} \cite{telea2004image} is another image inpainting algorithm based on fast marching method. 
	\end{itemize}
	
	\begin{table}[t]
		\centering
		\begin{tabular}{lccc}
			\hline
			& NLL & MAE (m) & RMSE (m) \\ \hline
			\bf SANP &  \bf -0.5538 & \bf 0.3096 & \bf 0.4439 \\ \hline
			SNP & -0.2147 & 0.41708 & 0.5793 \\ \hline
			ANP & -0.3775 & 0.3418 & 0.4865 \\ \hline
			NP & 0.8475 & 1.039 & 1.485  \\ \hline
			Linear & -  & 0.3586 & 0.5082 \\ \hline
			Cubic &  -  & 0.4034 & 0.6026 \\ \hline
			Nearest &  -  & 1.693 & 2.186\\ \hline
			Navier Stokes &  -  & 2.306 & 33.26\\ \hline
			Telea&  - & 9.131 & 44.06 \\ \hline
		\end{tabular}
		\caption{Reconstruction Results}
		\label{tab:recon}
	\end{table}

	\subsection{Reconstruction Results}
	Reconstruction errors of the SANP are compared against baseline models, in terms of three evaluation metrics: negative log-likelihood (NLL), mean absolute error (MAE), and root mean square error (RMSE).
	While MAE and RMSE only evaluate the accuracy of the reconstruction results, NLL further indicates how close the approximated \emph{probability distribution} is to the ground truth.
	Thus even if the evaluated MAE or RMSE is small, it can be inferred that if the NLL is large, the model is overconfident when predicting uncertain locations.
	As such, NLL represents the robustness of the model.
	Note that NLL is only reported for neural process models since others are not probabilistic models.
	
	The results are shown in in Table \ref{tab:recon}, SANP outperforms baseline algorithms for every metrics.
	From this result, we found that sparsification method helps the model to be more robust; this is consistent with previous studies that Dropout \cite{srivastava2014dropout} reduces the generalization error of the model in many machine learning algorithms.
	Moreover, the sparsification considerably improves the performance of NP; the sum aggregation in \eqref{eq:sum_agg} generally does not perform well in large window sizes because the latent information becomes vague as the number of data increased, but the sampling technique resolves such over-smoothing problem and enhances the accuracy of the information.
	
\begin{figure}[H]
		\begin{subfigure}{\textwidth}
			\centering
			\includegraphics[width=0.9\linewidth]{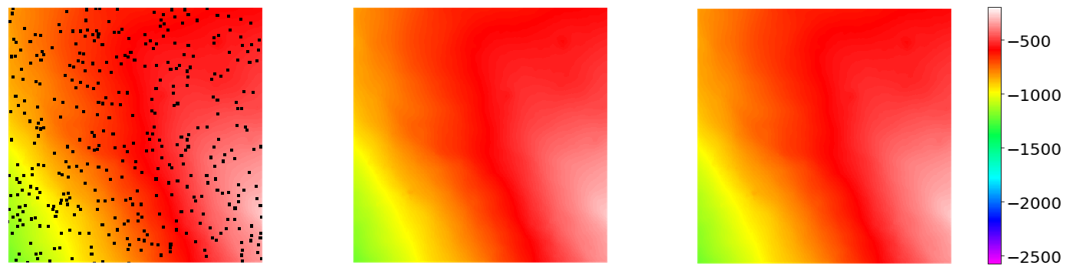}  
			\caption{}
		\end{subfigure}
		\begin{subfigure}{\textwidth}
			\centering
			\includegraphics[width=0.9\linewidth]{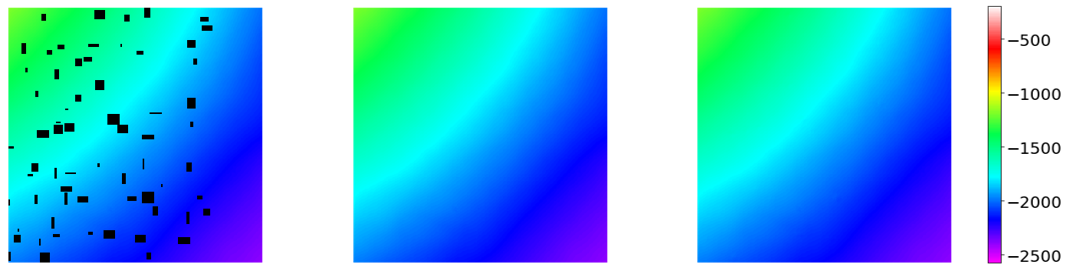}  
			\caption{}
		\end{subfigure}
		\begin{subfigure}{\textwidth}
			\centering
			\includegraphics[width=0.9\linewidth]{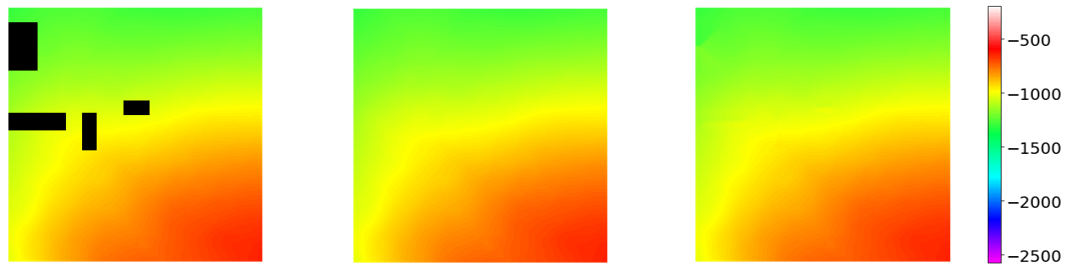}  
			\caption{}
		\end{subfigure}
		\begin{subfigure}{\textwidth}
			\centering
			\includegraphics[width=0.9\linewidth]{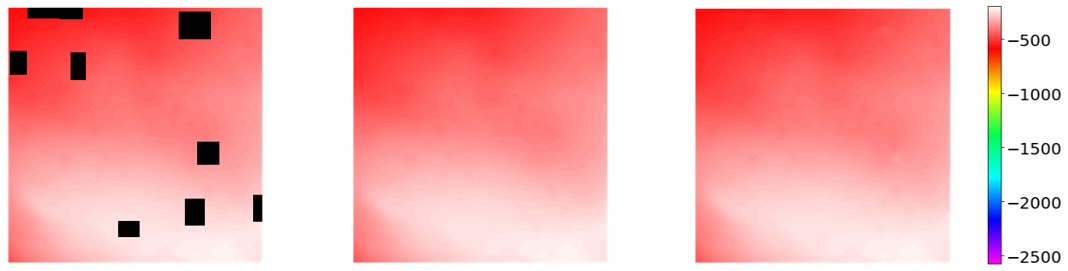}  
			\caption{}
		\end{subfigure}
		\begin{subfigure}{\textwidth}
			\centering
			\includegraphics[width=0.9\linewidth]{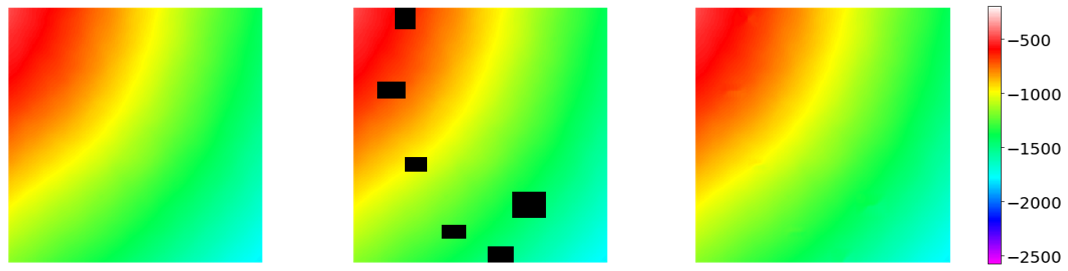}  
			\caption{}
		\end{subfigure}
		\caption{Sample reconstruction results of the SANP for various shapes and sizes of no-data gap. Color represents the elevation value (unit: $m$). (Left) input context, (middle) ground-truth, (right) reconstruction.}
		\label{fig:recon}
	\end{figure}

\begin{figure}[H]
		\begin{subfigure}{0.65\textwidth}
			%   \centering
			\includegraphics[width=\linewidth]{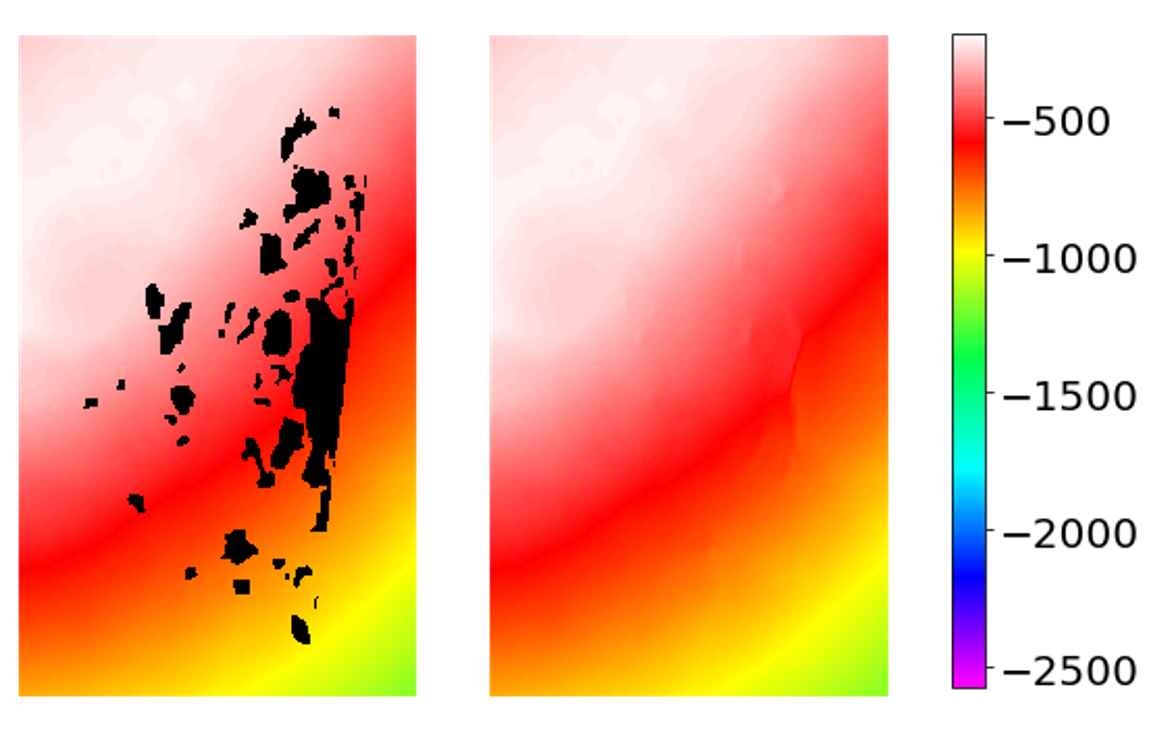}  
			\caption{Reconstructed map}
		\end{subfigure} \hfill
		\begin{subfigure}{0.3\textwidth}
			%   \centering
			\includegraphics[width=\linewidth]{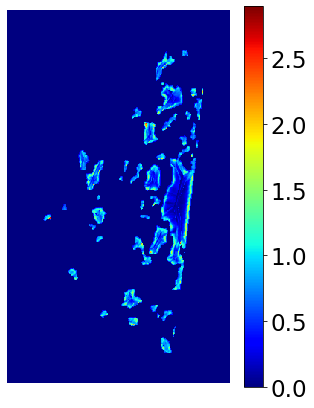}  
			\caption{Uncertainty map}
		\end{subfigure}
		\caption{Reconstruction results of the SANP on no-data gaps in NAC DEM data at the Apollo 17 landing site (20.0$^{\circ}$N and 30.4$^{\circ}$E). (a) Left image is NAC DEM, and right image is reconstructed map. Color represents the elevation value (unit: $m$). (b) Uncertainty map derived based on the standard deviation estimated by SANP. Color represents the standard deviation value (unit: $m$).}
		\label{fig:nodata}
	\end{figure}

	To illustrate more detailed results for our model, we primarily show several sample reconstruction results on artificially generated images with no-data gaps as shown in figure \ref{fig:recon}.
	As shown in the results, our model can predict the no-data gap successfully regardless of the size or shape of the voids.
	Furthermore, we provided the reconstruction result on NAC DEM at the Apollo 17 landing site which contains no-data gaps, and the uncertainty map estimated by SANP model.
	We fount that the SANP model predicts relatively high uncertainty where the reconstruction is suspected to the naked eye indicating that the model is aware of its own prediction accuracy, and it demonstrates the robustness of the model.
	Results are illustrated in figure \ref{fig:nodata}.
	
	\subsection{Ablation Studies} \label{sec:abl}
	To find the effect of hyperparameters to the model performance, ablation studies for SANP model are conducted in terms of the three most important factors: sampling size, sampling temperature, and latent dimension.
	In the case of sampling size and latent dimension, relative inference time is also compared as it is a factor directly related to the scalability of the method.
	
\begin{figure}[t]
		\begin{subfigure}{.45\textwidth}
			\centering
			% include first image
			\includegraphics[width=\linewidth]{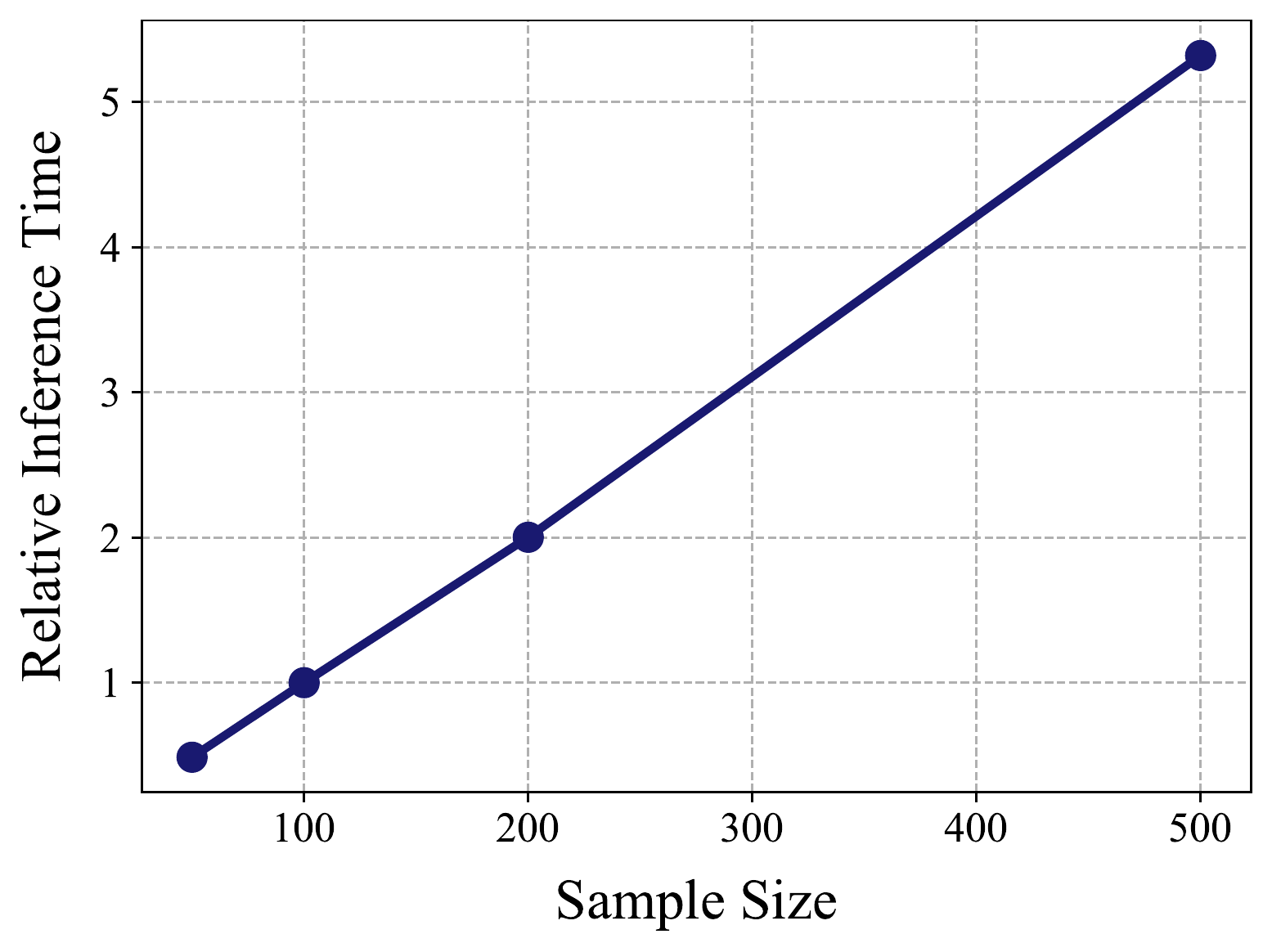}  
			\caption{}
			\label{fig:ss}
		\end{subfigure}
		\begin{subfigure}{.45\textwidth}
			\centering
			% include second image
			\includegraphics[width=\linewidth]{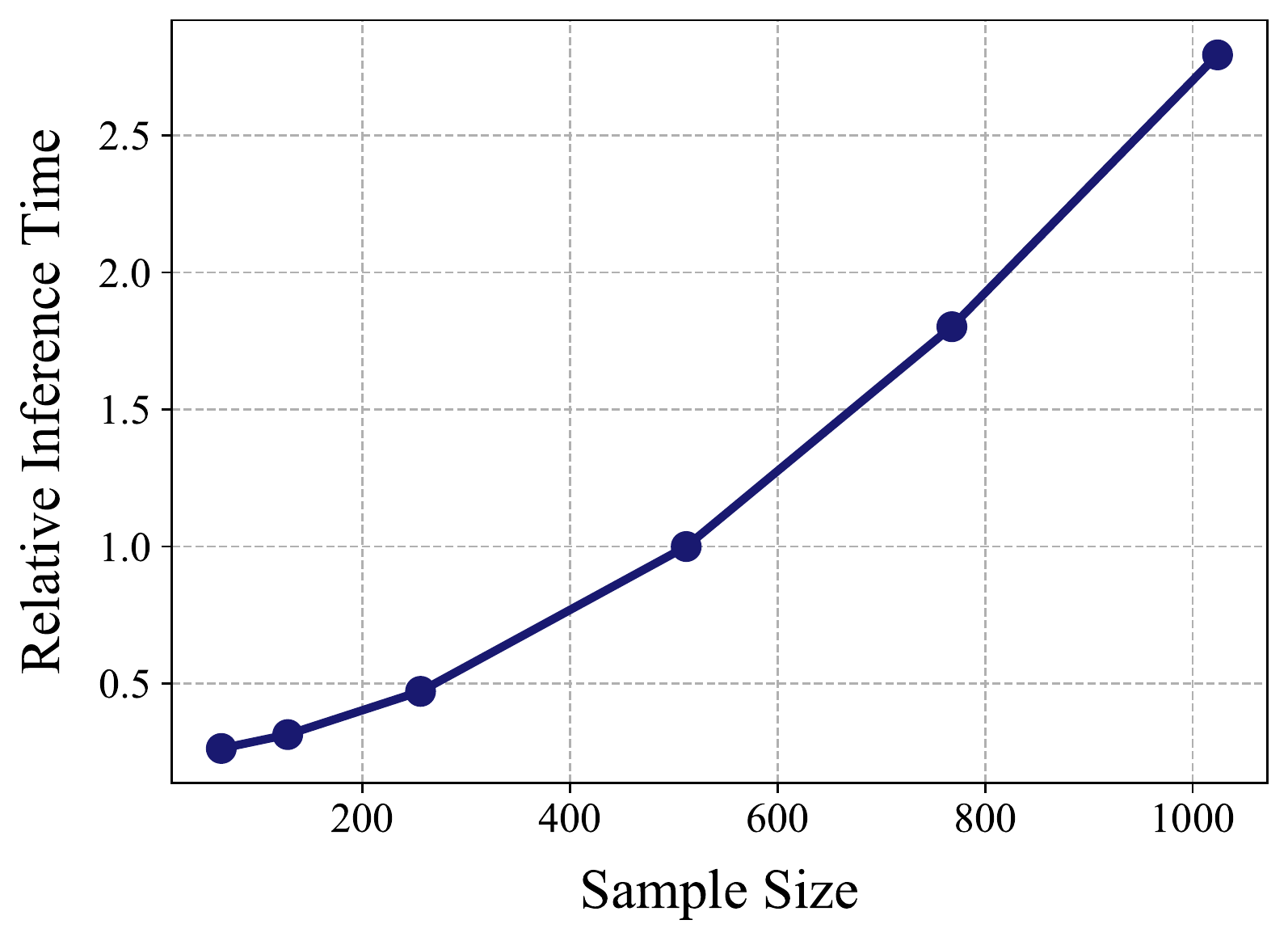}  
			\caption{}
			\label{fig:ed}
		\end{subfigure}
		\caption{The relative inference time by (a) sampling size and (b) latent dimension.}
	\end{figure}

	Primarily, the best sampling size is found to be 100 as shown in Table \ref{tab:ss}.
	A small number of context points can not provide enough information for reconstruction while too large number raises an over-smoothing problem and dramatically reduces the reconstruction quality.
	Empirically, it is found that over 1000 sample size leads to the out of memory (OOM) error for a single GPU with the batch size 1024. Figure \ref{fig:ss} shows that the inference time linearly increases to the number of sample size, as expected.
	
	Secondly, experiments on the effect of varying sampling temperature are conducted.
	Intuitively, it can be easily expected that the elevation values at points closer from the target point are more important.
	As such, the model with too high temperature would fail to make accurate reconstruction results.
	When the temperature is too low, meanwhile, the model becomes too myopic.
	As a result, the best sampling temperature is found at 0.4, as shown in Table \ref{tab:st}.
	
	Finally, the performance and inference time with different latent dimension is analyzed as well.
	As the latent dimension gets larger, the number of learnable parameters in neural networks gets bigger; therefore, the model with larger dimension generally has more representation power.
	However, the model with too large dimension may suffer from over-fitting problems, and the inference time quadratically increases with the latent dimension as well.
	In our experiments, 512 is found to be the most suitable dimension size as shown in Table \ref{tab:ed} and Figure \ref{fig:ed}.
	
	\begin{table}[t]
		\centering
		\begin{tabular}{l||ccc}
			\hline
			$K$ & NLL & MAE & RMSE \\ \hline
			50	& -0.3697	 & 0.3613	& 0.5131\\ \hline
			\bf 100	& \bf -0.5538	 & \bf 0.3096	& \bf 0.4439 \\ \hline
			200	& -0.1664	 & 0.3534	& 0.5018 \\ \hline
			500	& 0.5761	 & 0.8871	& 1.301	 \\ \hline
			1000	& \multicolumn{3}{c}{\it{Out of Memory}} \\ \hline
		\end{tabular}
		\caption{Reconstruction Performance by Sampling Size}
		\label{tab:ss}
	\end{table}
	
	\begin{table}[t]
		\centering
		\begin{tabular}{l||ccc}
			\hline
			$\alpha$ & NLL & MAE & RMSE \\ \hline
			$\infty$ & 668.8	& 60.90	& 69.11  \\ \hline
			8     & 0.7698	& 0.9701	& 1.329  \\ \hline
			0.8	& -0.4188	& 0.3379	& 0.4859  \\ \hline
			\bf 0.4	& \bf -0.5538	& \bf 0.3096	& \bf 0.4434  \\ \hline
			0.16	& -0.4081	& 0.3391	& 0.4867  \\ \hline
			0.08	& -0.3961	& 0.3362	& 0.4804  \\ \hline
			0	& -0.3134	& 0.3702	& 0.5252  \\ \hline
		\end{tabular}
		\caption{Reconstruction Performance by Sampling Temperature}
		\label{tab:st}
	\end{table}	
	
	\begin{table}[t]
		\centering
		\begin{tabular}{l||ccc}
			\hline
			$D$ & NLL & MAE & RMSE  \\ \hline
			128	&-0.4328	& 0.3371	& 0.4810\\ \hline
			256	&-0.4388	& 0.3295	& 0.4720 \\ \hline
			\bf 512	&\bf 0.5538	& \bf 0.3096	& \bf 0.4439 \\ \hline
			768	&-0.5033	& 0.3170	& 0.4550 \\ \hline
			1024 &-0.3334	& 0.3512	& 0.4973 \\ \hline
		\end{tabular}
		\caption{Reconstruction Performance by Latent Dimension}
		\label{tab:ed}
	\end{table}

	%---------------------------------------------------------------------------------
	\section{Conclusion}
	In this paper, we have proposed a probabilistic reconstruction framework for no-data gaps in lunar digital elevation maps. 
	Our framework is built upon attentive neural processes, a state of the art stochastic process model, which enables the model not only to perform uncertainty analysis but to be robust.
	To take account for scalability issue, we have extended neural process to sparse attentive neural process (SANP).
	SANP reduces the computational complexity of NP models and enhances reconstruction performance by solving over-fitting and over-smoothing problems.
	We have evaluated our model on lunar NAC DEMs at the Apollo 17 landing site and showed our model outperforms competitive void-filling methods in terms of negative likelihood, mean absolute error, and root mean square error.
	Furthermore, we have conducted extensive ablation studies to provide the effect of critical hyperparameters on the model performance.
	In future, the proposed method can be extended by using CNN, a powerful deep learning model used in various computer vision tasks, with NPs.
	
	%---------------------------------------------------------------------------------
	\section*{References}
	
	\bibliography{ref.bib}

\begin{thebibliography}{10}
\expandafter\ifx\csname url\endcsname\relax
  \def\url#1{\texttt{#1}}\fi
\expandafter\ifx\csname urlprefix\endcsname\relax\def\urlprefix{URL }\fi
\expandafter\ifx\csname href\endcsname\relax
  \def\href#1#2{#2} \def\path#1{#1}\fi

\bibitem{chin2007lunar}
G.~Chin, S.~Brylow, M.~Foote, J.~Garvin, J.~Kasper, J.~Keller, M.~Litvak,
  I.~Mitrofanov, D.~Paige, K.~Raney, et~al., Lunar reconnaissance orbiter
  overview: The{\'a}instrument suite and mission, Space Science Reviews 129~(4)
  (2007) 391--419.

\bibitem{smith2010lunar}
D.~E. Smith, M.~T. Zuber, G.~B. Jackson, J.~F. Cavanaugh, G.~A. Neumann,
  H.~Riris, X.~Sun, R.~S. Zellar, C.~Coltharp, J.~Connelly, et~al., The lunar
  orbiter laser altimeter investigation on the lunar reconnaissance orbiter
  mission, Space science reviews 150~(1-4) (2010) 209--241.

\bibitem{robinson2010lunar_a}
M.~Robinson, E.~Eliason, H.~Hiesinger, B.~Jolliff, A.~McEwen, M.~Malin,
  M.~Ravine, P.~Thomas, E.~Turtle, E.~Bowman-Cisneros, Lunar reconnaissance
  orbiter camera: first results, in: European Planetary Science Congress 2010,
  2010, p. 457.

\bibitem{robinson2010lunar_b}
M.~Robinson, S.~Brylow, M.~Tschimmel, D.~Humm, S.~Lawrence, P.~Thomas,
  B.~Denevi, E.~Bowman-Cisneros, J.~Zerr, M.~Ravine, et~al., Lunar
  reconnaissance orbiter camera (lroc) instrument overview, Space science
  reviews 150~(1-4) (2010) 81--124.

\bibitem{tran2010generating}
T.~Tran, M.~Rosiek, R.~A. Beyer, S.~Mattson, E.~Howington-Kraus, M.~Robinson,
  B.~Archinal, K.~Edmundson, D.~Harbour, E.~Anderson, et~al., Generating
  digital terrain models using lroc nac images, in: At ASPRS/CaGIS 2010 Fall
  Specialty Conference, 2010.

\bibitem{yang2016neural}
Y.~Yang, Y.~Yan, Neural network approximation-based nonsingular terminal
  sliding mode control for trajectory tracking of robotic airships, Aerospace
  Science and Technology 54 (2016) 192--197.

\bibitem{chang2017adaptive}
Y.~Chang, T.~Jiang, Z.~Pu, Adaptive control of hypersonic vehicles based on
  characteristic models with fuzzy neural network estimators, Aerospace Science
  and Technology 68 (2017) 475--485.

\bibitem{bagherzadeh2018nonlinear}
S.~A. Bagherzadeh, Nonlinear aircraft system identification using artificial
  neural networks enhanced by empirical mode decomposition, Aerospace Science
  and Technology 75 (2018) 155--171.

\bibitem{furfaro2018deep}
R.~Furfaro, I.~Bloise, M.~Orlandelli, P.~Di~Lizia, F.~Topputo, R.~Linares,
  et~al., Deep learning for autonomous lunar landing, in: 2018 AAS/AIAA
  Astrodynamics Specialist Conference, 2018, pp. 1--22.

\bibitem{roy2019lunar}
H.~Roy, S.~Chaudhury, T.~Yamasaki, D.~DeLatte, M.~Ohtake, T.~Hashimoto, Lunar
  surface image restoration using u-net based deep neural networks, in: Lunar
  and Planetary Science Conference, Vol.~50, 2019.

\bibitem{reuter2007evaluation}
H.~I. Reuter, A.~Nelson, A.~Jarvis, An evaluation of void-filling interpolation
  methods for srtm data, International Journal of Geographical Information
  Science 21~(9) (2007) 983--1008.

\bibitem{jassim2013image}
F.~A. Jassim, F.~H. Altaany, Image interpolation using kriging technique for
  spatial data, arXiv preprint arXiv:1302.1294.

\bibitem{bertalmio2000image}
M.~Bertalmio, G.~Sapiro, V.~Caselles, C.~Ballester, Image inpainting, in:
  Proceedings of the 27th annual conference on Computer graphics and
  interactive techniques, ACM Press/Addison-Wesley Publishing Co., 2000, pp.
  417--424.

\bibitem{ballester2000filling}
C.~Ballester, M.~Bertalmio, V.~Caselles, G.~Sapiro, J.~Verdera, Filling-in by
  joint interpolation of vector fields and gray levels.

\bibitem{efros2001image}
A.~A. Efros, W.~T. Freeman, Image quilting for texture synthesis and transfer,
  in: Proceedings of the 28th annual conference on Computer graphics and
  interactive techniques, ACM, 2001, pp. 341--346.

\bibitem{barnes2009patchmatch}
C.~Barnes, E.~Shechtman, A.~Finkelstein, D.~B. Goldman, Patchmatch: A
  randomized correspondence algorithm for structural image editing, in: ACM
  Transactions on Graphics (ToG), Vol.~28, ACM, 2009, p.~24.

\bibitem{komodakis2006image}
N.~Komodakis, Image completion using global optimization, in: 2006 IEEE
  Computer Society Conference on Computer Vision and Pattern Recognition
  (CVPR'06), Vol.~1, IEEE, 2006, pp. 442--452.

\bibitem{krizhevsky2012imagenet}
A.~Krizhevsky, I.~Sutskever, G.~E. Hinton, Imagenet classification with deep
  convolutional neural networks, in: Advances in neural information processing
  systems, 2012, pp. 1097--1105.

\bibitem{goodfellow2014generative}
I.~Goodfellow, J.~Pouget-Abadie, M.~Mirza, B.~Xu, D.~Warde-Farley, S.~Ozair,
  A.~Courville, Y.~Bengio, Generative adversarial nets, in: Advances in neural
  information processing systems, 2014, pp. 2672--2680.

\bibitem{arjovsky2017wasserstein}
M.~Arjovsky, S.~Chintala, L.~Bottou, Wasserstein gan, arXiv preprint
  arXiv:1701.07875.

\bibitem{iizuka2017globally}
S.~Iizuka, E.~Simo-Serra, H.~Ishikawa, Globally and locally consistent image
  completion, ACM Transactions on Graphics (ToG) 36~(4) (2017) 107.

\bibitem{yeh2017semantic}
R.~A. Yeh, C.~Chen, T.~Yian~Lim, A.~G. Schwing, M.~Hasegawa-Johnson, M.~N. Do,
  Semantic image inpainting with deep generative models, in: Proceedings of the
  IEEE Conference on Computer Vision and Pattern Recognition, 2017, pp.
  5485--5493.

\bibitem{yang2017high}
C.~Yang, X.~Lu, Z.~Lin, E.~Shechtman, O.~Wang, H.~Li, High-resolution image
  inpainting using multi-scale neural patch synthesis, in: Proceedings of the
  IEEE Conference on Computer Vision and Pattern Recognition, 2017, pp.
  6721--6729.

\bibitem{gavriil2019void}
K.~Gavriil, G.~Muntingh, O.~J. Barrowclough, Void filling of digital elevation
  models with deep generative models, IEEE Geoscience and Remote Sensing
  Letters.

\bibitem{vaswani2017attention}
A.~Vaswani, N.~Shazeer, N.~Parmar, J.~Uszkoreit, L.~Jones, A.~N. Gomez,
  {\L}.~Kaiser, I.~Polosukhin, Attention is all you need, in: Advances in
  neural information processing systems, 2017, pp. 5998--6008.

\bibitem{garnelo2018neural}
M.~Garnelo, J.~Schwarz, D.~Rosenbaum, F.~Viola, D.~J. Rezende, S.~Eslami, Y.~W.
  Teh, Neural processes, arXiv preprint arXiv:1807.01622.

\bibitem{garnelo2018conditional}
M.~Garnelo, D.~Rosenbaum, C.~J. Maddison, T.~Ramalho, D.~Saxton, M.~Shanahan,
  Y.~W. Teh, D.~J. Rezende, S.~Eslami, Conditional neural processes, arXiv
  preprint arXiv:1807.01613.

\bibitem{kim2019attentive}
H.~Kim, A.~Mnih, J.~Schwarz, M.~Garnelo, A.~Eslami, D.~Rosenbaum, O.~Vinyals,
  Y.~W. Teh, Attentive neural processes, arXiv preprint arXiv:1901.05761.

\bibitem{rasmussen2006gaussian}
C.~E. Rasmussen, C.~K. Williams, Gaussian processes for machine learning,
  Vol.~1, MIT press Cambridge, 2006.

\bibitem{devlin2018bert}
J.~Devlin, M.-W. Chang, K.~Lee, K.~Toutanova, Bert: Pre-training of deep
  bidirectional transformers for language understanding, arXiv preprint
  arXiv:1810.04805.

\bibitem{yu2018generative}
J.~Yu, Z.~Lin, J.~Yang, X.~Shen, X.~Lu, T.~S. Huang, Generative image
  inpainting with contextual attention, in: Proceedings of the IEEE Conference
  on Computer Vision and Pattern Recognition, 2018, pp. 5505--5514.

\bibitem{zhou2018deep}
G.~Zhou, X.~Zhu, C.~Song, Y.~Fan, H.~Zhu, X.~Ma, Y.~Yan, J.~Jin, H.~Li, K.~Gai,
  Deep interest network for click-through rate prediction, in: Proceedings of
  the 24th ACM SIGKDD International Conference on Knowledge Discovery \& Data
  Mining, ACM, 2018, pp. 1059--1068.

\bibitem{kingma2013auto}
D.~P. Kingma, M.~Welling, Auto-encoding variational bayes, arXiv preprint
  arXiv:1312.6114.

\bibitem{gershman2014amortized}
S.~Gershman, N.~Goodman, Amortized inference in probabilistic reasoning, in:
  Proceedings of the annual meeting of the cognitive science society, Vol.~36,
  2014.

\bibitem{kingma2014adam}
D.~Kingma, J.~Ba, Adam: A method for stochastic optimization, arXiv preprint
  arXiv:1412.6980.

\bibitem{NEURIPS2019_9015}
A.~Paszke, S.~Gross, F.~Massa, A.~Lerer, J.~Bradbury, G.~Chanan, T.~Killeen,
  Z.~Lin, N.~Gimelshein, L.~Antiga, A.~Desmaison, A.~Kopf, E.~Yang, Z.~DeVito,
  M.~Raison, A.~Tejani, S.~Chilamkurthy, B.~Steiner, L.~Fang, J.~Bai,
  S.~Chintala,
  \href{http://papers.neurips.cc/paper/9015-pytorch-an-imperative-style-high-performance-deep-learning-library.pdf}{Pytorch:
  An imperative style, high-performance deep learning library}, in: H.~Wallach,
  H.~Larochelle, A.~Beygelzimer, F.~d\textquotesingle Alch\'{e}-Buc, E.~Fox,
  R.~Garnett (Eds.), Advances in Neural Information Processing Systems 32,
  Curran Associates, Inc., 2019, pp. 8024--8035.
\newline\urlprefix\url{http://papers.neurips.cc/paper/9015-pytorch-an-imperative-style-high-performance-deep-learning-library.pdf}

\bibitem{jones2001scipy}
E.~Jones, T.~Oliphant, P.~Peterson, et~al., Scipy: Open source scientific tools
  for python.

\bibitem{opencv_library}
G.~Bradski, {The OpenCV Library}, Dr. Dobb's Journal of Software Tools.

\bibitem{bertalmio2001navier}
M.~Bertalmio, A.~L. Bertozzi, G.~Sapiro, Navier-stokes, fluid dynamics, and
  image and video inpainting, in: Proceedings of the 2001 IEEE Computer Society
  Conference on Computer Vision and Pattern Recognition. CVPR 2001, Vol.~1,
  IEEE, 2001, pp. I--I.

\bibitem{telea2004image}
A.~Telea, An image inpainting technique based on the fast marching method,
  Journal of graphics tools 9~(1) (2004) 23--34.

\bibitem{srivastava2014dropout}
N.~Srivastava, G.~Hinton, A.~Krizhevsky, I.~Sutskever, R.~Salakhutdinov,
  Dropout: a simple way to prevent neural networks from overfitting, The
  journal of machine learning research 15~(1) (2014) 1929--1958.

\end{thebibliography}
	
\end{document}